\documentclass[dvips]{article}

\usepackage{icrc2011}

\title{Observation of the BL Lac objects 1ES\,1215+303 and 1ES\,1218+304 with the MAGIC telescopes}

\newcommand{\etal}{\MakeLowercase{\textit{et al. }}} 
\shorttitle{colin \etal Observation of 1ES\,1215+303 and 1ES\,1218+304 with MAGIC}

\authors{Pierre Colin$^{1}$,
 Josefa Becerra Gonz\'alez$^{2}$,
 Elina Lindfors$^{3}$,
 Saverio Lombardi$^{4}$,
 Julian Sitarek$^{5}$ on behalf of the MAGIC collaboration
}

\afiliations{$^1$Max-Planck-Institut für Physik, Munich, Germany\\
 $^2$Instituto de Astrofisica de Canarias, La Laguna, Spain\\
 $^3$Tuorla Observatory, Piikkiö, Finland\\
 $^4$Universit\`{a} di Padova and INFN, I-35131 Padova, Italy\\
 $^5$University of Lodz, Poland }
\email{colin@mppmu.mpg.de}

\abstract{The two BL Lac objects 1ES\,1215+303 and 1ES\,1218+304, separated by 0.8$^\circ$, were
observed with the MAGIC telescopes in 2010 and 2011. The 20 hours of data registered
in January 2011 resulted in the first detection at Very High Energy ($>$100\,GeV) of 1ES\,1215+303
(also known as ON-325). This observation was triggered by a high optical state of the source reported
by the Tuorla blazar monitoring program. Comparison with the 25 hours of data carried out from January
to May 2010 suggests that 1ES\,1215+303 was flaring also in VHE gamma-rays in 2011.
In addition, the Swift ToO observations in X-rays showed that the flux was almost doubled respect
to previous observations (December 2009). Instead, 1ES\,1218+304 is a well known VHE gamma-ray emitter lying in
the same field of view, which was then simultaneously observed with the MAGIC telescopes.
The overall observation time of nearly 45 hours has permitted to measure
the spectrum of this source with a much higher precision than previously reported by
MAGIC. Here, we present the results of the MAGIC and the multi-wavelength observations of
these two VHE gamma-ray emitting AGNs.}

\begin{document}
\maketitle

\section{Introduction}

BL Lac objects are a special type of active galactic nuclei (AGN) with broad-band emission from radio waves to $\gamma$-rays dominated by non-thermal emission without (or with faint) lines. They show strong and rapid variability at all wavelengths and 
have a spectral energy distribution (SED) with a typical double bump shape. The emission is understood as originating from the relativistic electrons of the AGN jet which is pointing very close to our line of sight. The first and second bumps are associated respectively to synchrotron and inverse Compton emissions.

Most of the extragalactic Very High Energy (VHE, $>$\,100\,GeV) $\gamma$-ray sources are BL Lac objects and more specifically High-frequency-peaked BL Lac (HBL) with the first SED bump peaking in the X-ray band. The second bump of the intrinsic SED
of HBL should peak in the GeV-TeV $\gamma$-ray regime. However the VHE emission is attenuated by its interaction with the extragalactic background light (EBL) during its travel toward Earth and only nearby HBL (redshift$<$0.5) can be observed at VHE. In order to obtain the intrinsic VHE spectrum, the measured spectrum has to be corrected according to an EBL model and to the distance of the source.
Constraints on the EBL models can also be derived from the measured VHE spectra.

1ES\,1215+303 (also known as ON\,325) is a HBL with an uncertain redshift (two values can be found in the literature:
z\,=\,0.130 and z\,=\,0.237). The source was classified as promising candidate TeV blazar by Costamante \& Ghisellini \cite{Cos02}
and has been observed several times in VHE $\gamma$-rays prior to the observations presented here.
The Whipple (10\,m telescope) and MAGIC (single telescope observation) reported respectively the following integral flux upper limits:
F$_{>430GeV}$$<$1.89$\times10^{-11}$\,cm$^{-2}$\,s$^{-1}$ in 2000 \cite{Hor04} and 
F$_{>120GeV}$$<$3.5$\times10^{-11}$\,cm$^{-2}$\,s$^{-1}$ in 2007-2008 \cite{Ale11}.


1ES\,1218+304 is another HBL located only 0.8$^\circ$ away from 1ES\,1215+303.
It has a moderate redshift of 0.182. It was first detected at VHE by MAGIC \cite{Alb06}
and confirmed by VERITAS \cite{Acc09} who reported also fast variability from this source in 2009 \cite{Acc10}.
In the \textit{Fermi}-LAT one year catalog \cite{Abd10b}, the source is flagged non-variable.
It is an interesting object because the measured VHE spectrum is particularly hard for this redshift, indicating 
an intrinsic SED with an inverse Compton peak above 1\,TeV. The emission is then strongly interacting with the EBL
before reaching us. This makes it a good candidate to probe the EBL \cite{Alb06,Acc09} or the extragalactic magnetic field \cite{Ner10}.

This proceeding reports on the 2010 and 2011 measurements of the VHE $\gamma$-ray emission from
both 1ES\,1215+303 and 1ES\,1218+304 with the MAGIC telescopes. We also compare the VHE emissions with the longterm
optical light curves obtained with the Tuorla blazar monitoring program (KVA telescope).


\section{Observations and Data Analysis}

\subsection{MAGIC}
MAGIC consists of two 17\,m imaging air Cherenkov telescopes located at the Canary Island of La Palma, 2200\,m above sea level. The stereoscopic system has been in operation since fall 2009 and reaches its best sensitivity above $\sim$250\,GeV with 0.8\% of the Crab Nabula in 50\,h \cite{Car11} . MAGIC cameras have a field of view of 3.5$^\circ$.

1ES\,1215+303 and 1ES\,1218+304 were observed with the MAGIC telescopes in January-February 2010, May-June 2010 and January-February 2011 for a total of 48\,h. The observations were done in wobble mode around 1ES\,1215+303 with four pointing positions 0.4$^\circ$ away from this source. 1ES\,1218+304 was not the primary target of these observations, but it is always inside the MAGIC camera field of view. It lies at different distances from the camera center in each pointing position: 0.36$^\circ$, 0.84$^\circ$, 0.87$^\circ$ and 1.15$^\circ$. Hence, 1ES\,1218+304 can be studied even if the MAGIC performance degrades at large offset \cite{Car11}.

The data were taken during dark night and moderate moon conditions at zenith angles from 1$^\circ$ to 40$^\circ$.
For the analysis the data were divided into two samples corresponding to two observing epochs: 2010 ($\sim$26\,h) and 2011 ($\sim$22\,h).
The data were analyzed using standard MAGIC software \cite{Mor09} with additional
adaptations incorporating the stereoscopic observations \cite{Lom11}.
Due to the presence of two sources in the same field of view, a special care has to be taken to estimate the background.

\subsection{Tuorla blazar monitoring program}

The 2011 MAGIC observations were triggered by an optical high-state of 1ES\,1215+303
reported by the Tuorla blazar monitoring program \cite{tbmp}.

Since the beginning of its scientific operation, MAGIC has been successfully performing optically
triggered Target of Opportunity (ToO) observations of AGNs .
The triggers have been provided by the Tuorla blazar monitoring program
using the 35\,cm KVA optical telescope. The KVA telescope is located on La Palma, but is operated remotely from Finland. The observations are done in R-band and the R-band magnitude of the source is measured from CCD images using differential photometry.

1ES\,1215+303 and 1ES\,1218+304 have been part of the Tuorla blazar monitoring program from its beginning and then have been observed regularly since 2002. The core flux of the sources (shown Figure~\ref{KVA_LC}) is measured by subtracting the host galaxy contribution from the observed flux \cite{Nil07}.

\section{Results}
All the results presented here are preliminary and may evolve before the conference.

\subsection{1ES\,1215+303}
\begin{figure}[!t]
  \vspace{5mm}
  \centering
  \includegraphics[width=3.in]{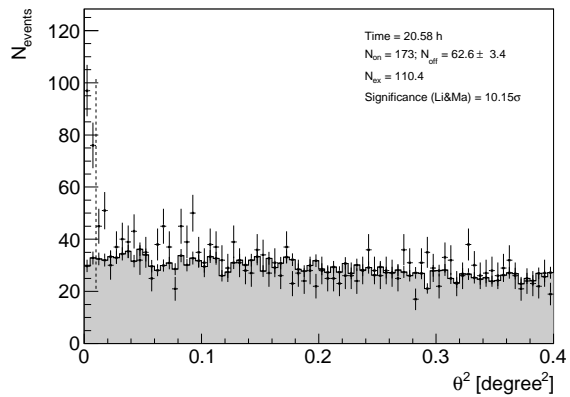}
  \caption{Distribution of the square distances ($\theta^2$) between the reconstructed shower directions
and the 1ES\,1215+303 position, for data taken in January-February 2011 with MAGIC.
The grey filled histogram represents the expected background estimated with 3 Off positions
at the same distance from the camera center.}
  \label{theta2_2011}
 \end{figure}

The analysis of the 2010 data resulted in 3\,$\sigma$\footnote{$\sigma$ = standard deviation}
excess from 1ES\,1215+303 while for 2011 data
an excess corresponding to 10.1\,$\sigma$ has been detected. Figure~\ref{theta2_2011} shows the distribution
of the square distance ($\theta^2$) between the reconstructed shower directions and 1ES\,1215+303
where a clear excess appears at small $\theta^2$. The energy threshold of this analysis is about 200\,GeV.
The 2011 $\theta^2$ plot represents the first significant detection of VHE $\gamma$-rays from this source \cite{Mar11}.

\begin{figure}[!t]
  \vspace{5mm}
  \centering
  \includegraphics[width=3.in]{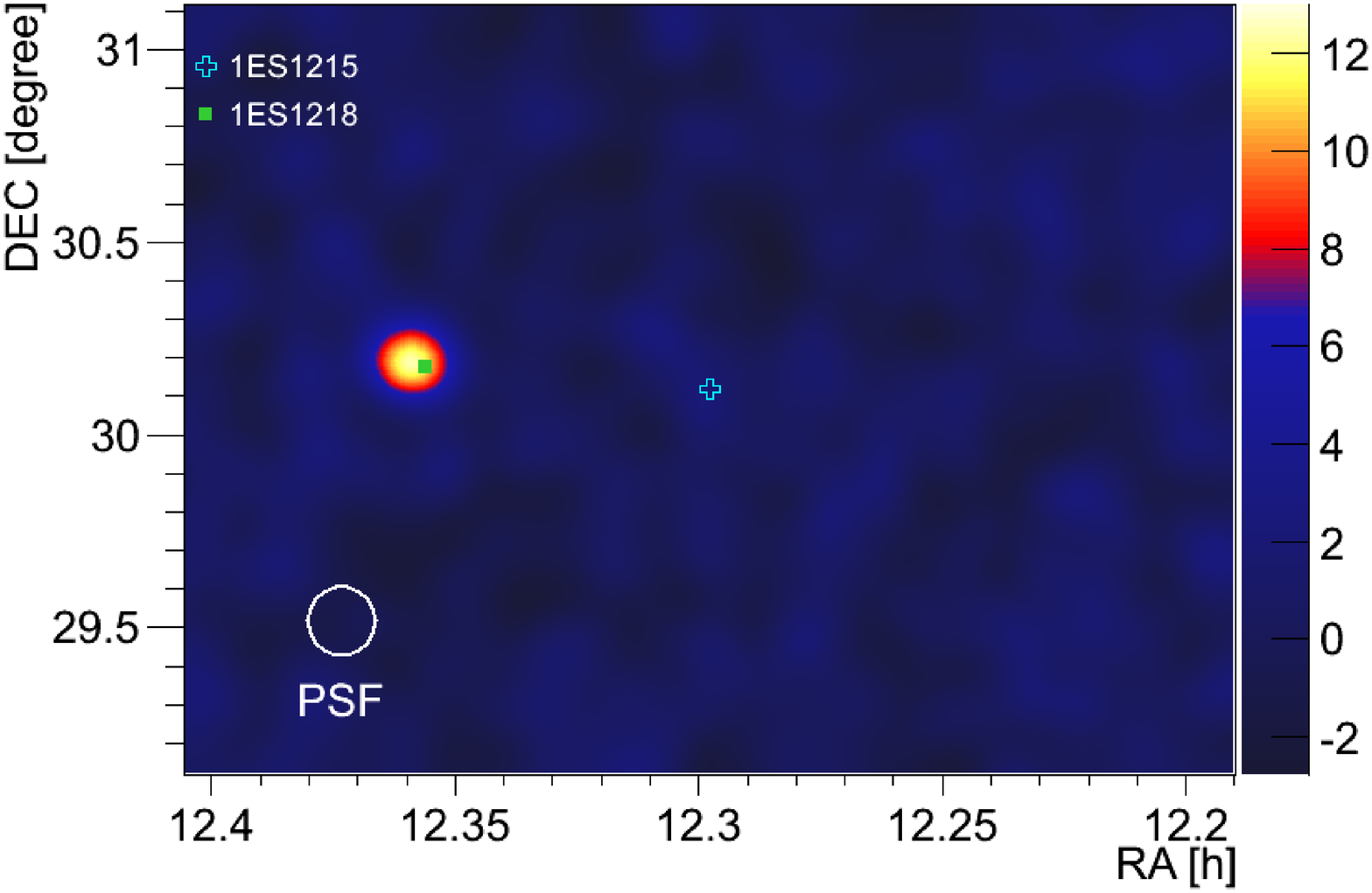}
  \includegraphics[width=3.in]{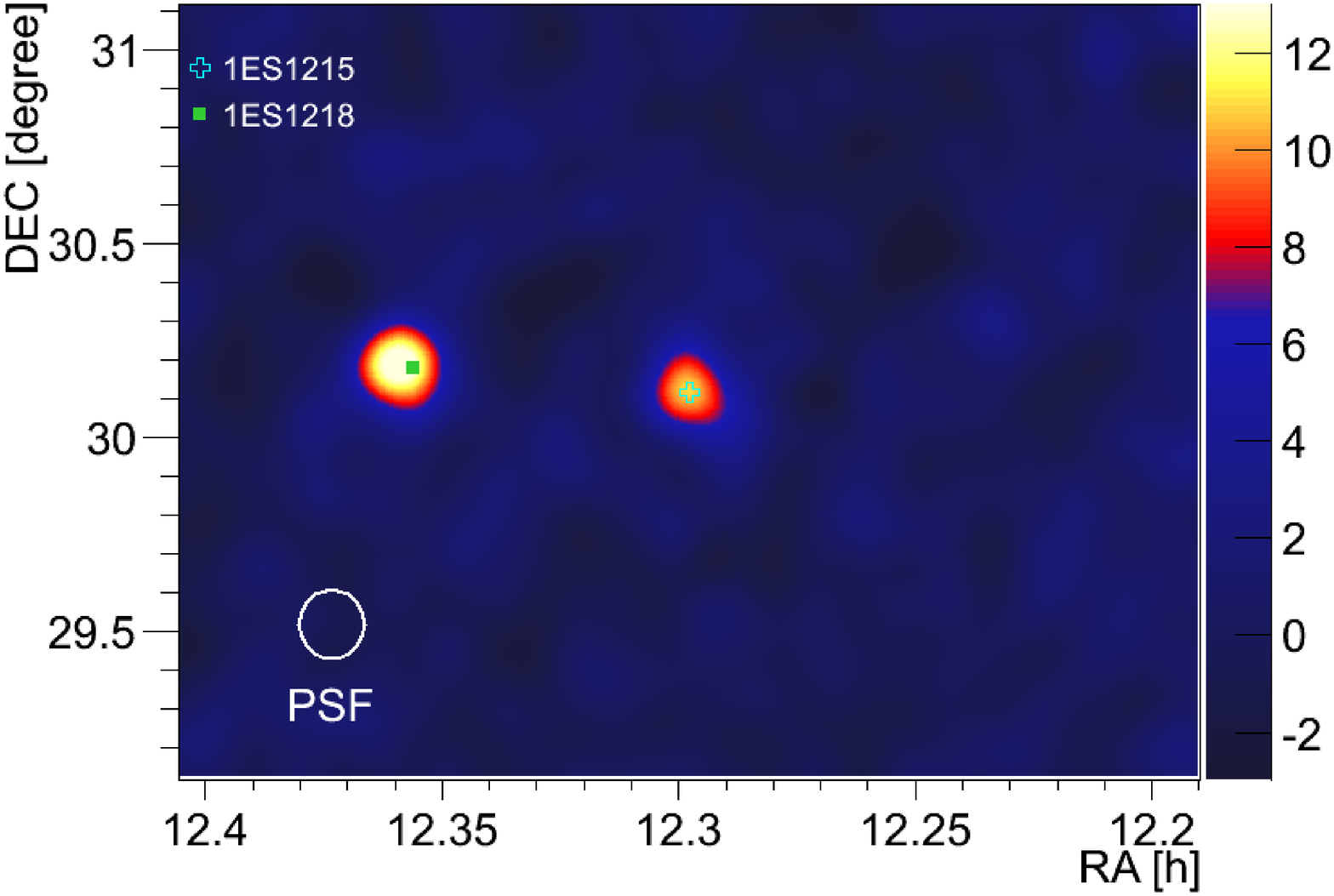}
  \caption{Significance skymaps corresponding to 2010 (top) and 2011 (bottom) observations.}
  \label{skymaps}
 \end{figure}

The significance skymaps of a point-like source detection inside the observed field of view are shown 
Figure~\ref{skymaps} for both 2010 and 2011 data samples. These two samples contain about the same amount of data and both maps
have alomost the same sensitivity. 1ES\,1215+303 is clearly visible at the center of the map in 2011
but did not show up in 2010, suggesting year scale variability.

\begin{figure*}[!t]
   \centerline{\includegraphics[width=3.in]{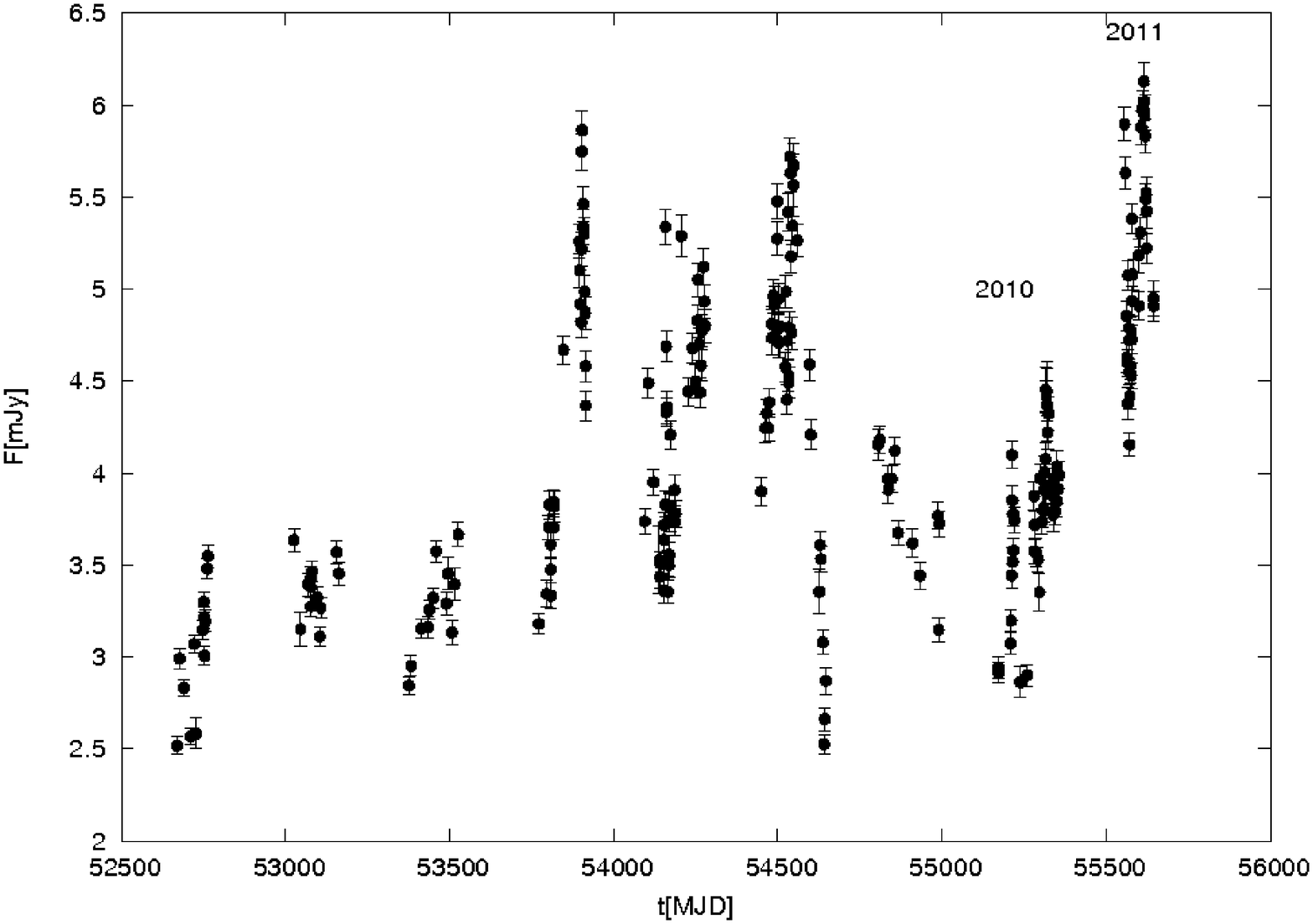}\label{fig4}
              \hfil
              \includegraphics[width=3.in]{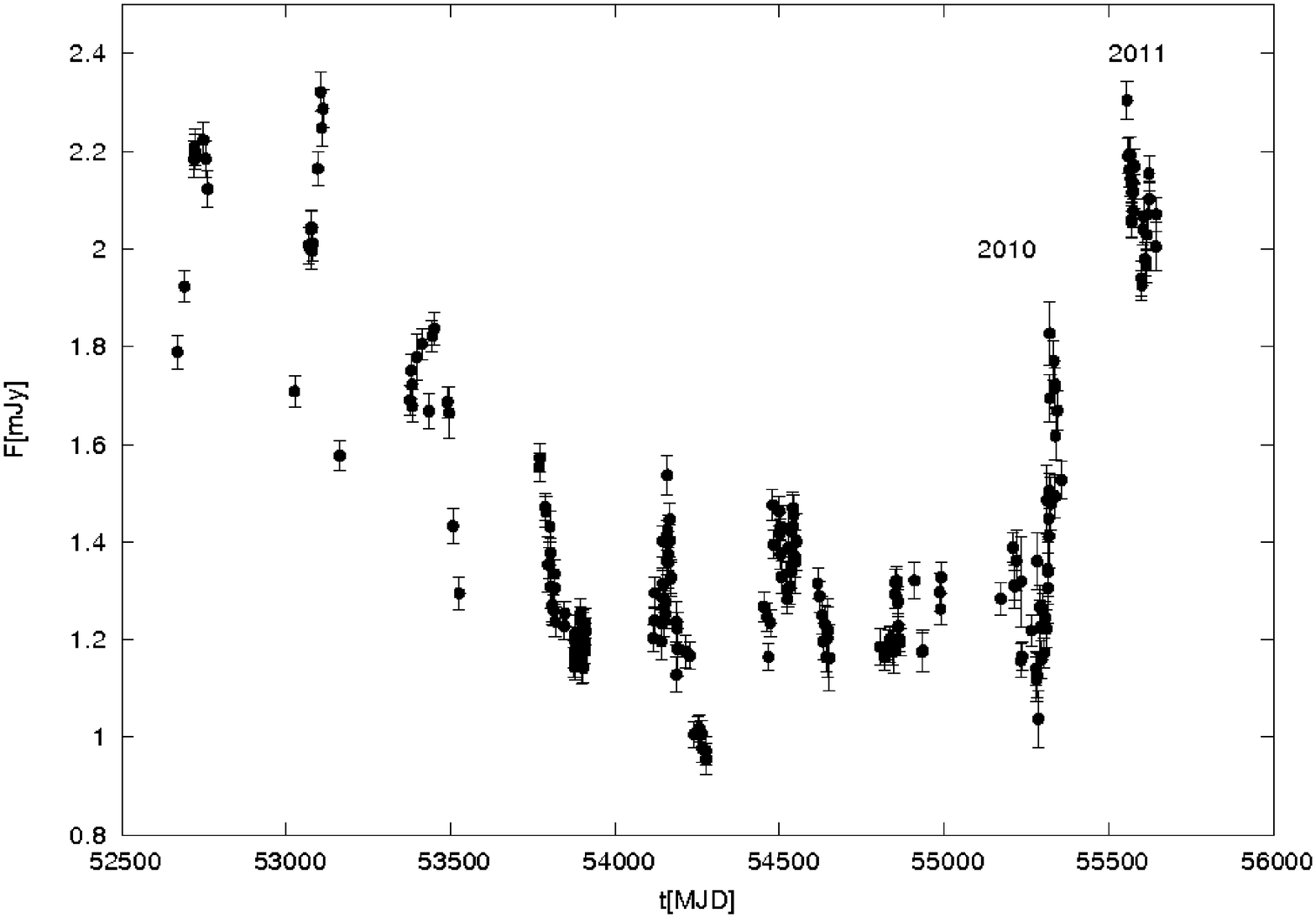} \label{fig5}
             }
   \caption{Longterm optical lightcurves of 1ES 1215+303 (left) and 1ES 1218+304
(right) from the Tuorla blazar monitoring program. Both sources were in a high state
in optical band during the 2011 observations and in a lower state in 2010.}
   \label{KVA_LC}
 \end{figure*}

The optical state of 1ES\,1215+303 was also higher in 2011 than in 2010. The core flux measured with the KVA 
telescope was about a factor 2 higher (see the lightcurves in Figure~\ref{KVA_LC}).

1ES\,1215+303 is the fifth BL Lac objects discovered by MAGIC during an optical high state \cite{Alb06b,Alb07,And09,Mar10,Mar11}.
This strongly suggests there is a connection between optical and VHE $\gamma$-ray high states in BL Lac objects. However, follow-up observations of the discovered sources have not been able to confirm such a connection \cite{Rug11, Rei11}.

\subsection{1ES\,1218+304}
The preliminary analysis of 1ES\,1218+304 results in a clear detection of the source with a significance of $\sim$20\,$\sigma$ for the full data set.
As one can see in Figure~\ref{skymaps}, the source shows approximately the same significances in both maps suggesting a similar flux in 2010 and 2011. On the contrary, the optical emission reported in Figure~\ref{KVA_LC} increases of about a factor 2 between 2010 and 2011. The possible connection between optical and VHE emissions seen in 1ES\,1215+303 seems not to be present in 1ES\,1218+304 data.

The lightcurve and spectrum of 1ES\,1218+304 are still under study and will be presented at the conference.

\section{Conclusion and prospects}
The observation of the HBL 1ES\,1215+303 with the MAGIC telescopes in 2010 and 2011
resulted in the discovery of VHE signal from this source (in 2011) and in the deep observation
of the well-know neighboring TeV HBL 1ES\,1218+304.

Concerning 1ES\,1215+303, multiwavelength data simultaneous and quasi-simultaneous
to MAGIC observations were collected from radio to the $\gamma$-ray regime including
e.g. Mets\"ahovi 37\,GHz data, BVR optical data, optical R-band (presented here) and polarization data from KVA,
Swift and \textit{Fermi}-LAT data. The analysis and interpretation of these data are still ongoing.

For 1ES 1218+304 the simultaneous spectra from \textit{Fermi}-LAT and MAGIC cover continuously more
than 3 orders of magnitude in energy and could bring new constraints on the EBL 
and on intergalactic magnetic field (the study is ongoing).

\section{Acknowledgement}

We would like to thank the Instituto de Astrof\'{\i}sica de
Canarias for the excellent working conditions at the
Observatorio del Roque de los Muchachos in La Palma.
The support of the German BMBF and MPG, the Italian INFN, 
the Swiss National Fund SNF, and the Spanish MICINN is 
gratefully acknowledged. This work was also supported by 
the Marie Curie program, by the CPAN CSD2007-00042 and MultiDark
CSD2009-00064 projects of the Spanish Consolider-Ingenio 2010
programme, by grant DO02-353 of the Bulgarian NSF, by grant 127740 of 
the Academy of Finland, by the YIP of the Helmholtz Gemeinschaft, 
by the DFG Cluster of Excellence ``Origin and Structure of the 
Universe'', by the DFG Collaborative Research Centers SFB823/C4 and SFB876/C3,
and by the Polish MNiSzW grant 745/N-HESS-MAGIC/2010/0.

\clearpage

\end{document}